\documentclass[showpacs,amsmath,amssymb]{revtex4}

\usepackage{graphicx}
\usepackage{dcolumn}
\usepackage{bm}
\usepackage{epsfig}

\begin{document}

\def\0#1#2{\frac{#1}{#2}}
\def\bct{\begin{center}} \def\ect{\end{center}}
\def\beq{\begin{equation}} \def\eeq{\end{equation}}
\def\bea{\begin{eqnarray}} \def\eea{\end{eqnarray}}
\def\nnu{\nonumber}
\def\n{\noindent} \def\pl{\partial}
\def\g{\gamma}  \def\O{\Omega} \def\e{\varepsilon} \def\o{\omega}
\def\s{\sigma}  \def\b{\beta} \def\p{\psi} \def\r{\rho}
\def\G{\Gamma} \def\S{\Sigma}

\title{thermodynamical consistent CJT calculation in studying nuclear matter}
\author{Song~Shu}
\author{Jia-Rong~Li}
\affiliation{Institute of Particle Physics, Hua-Zhong Normal
University, Wuhan 430079, P. R. China}
\begin{abstract}
We have attempted to apply the CJT formalism to study the nuclear
matter. The thermodynamic potential is calculated in Hartree-Fock
approximation in the CJT formalism. After neglecting the medium
effects to the mesons, the numerical results are found very
consistent with those obtained from the mean field calculation. In
our calculation the thermodynamical consistency is also preserved.
\end{abstract}  \maketitle

\section{Introduction}
Mean field theory (MFT) of quantum hadrodynamics (QHD) is very
successful in describing many nuclear phenomena~\cite{a,b,b1}. The
idea of MFT, in a simple description, is that at high enough
nuclear density, fluctuations in the meson fields could be
ignored, and the meson fields could be replaced by their classical
expectation values or mean fields~\cite{b2}. However, mean fields
are insufficient for a detail understanding of short-distance
nuclear physics. The development of reliable techniques to go
beyond mean field calculation is also required~\cite{b}. In a
renormalizable field theory, through a Green's function formalism,
one can get relativistic Hartree approximation (RHA) by
self-consistently summing all the tadpole graphs in the baryon
self-energy~\cite{c,d}. It turns out that RHA consists MFT result
and the additional vacuum fluctuation corrections. Neglecting the
vacuum corrections, RHA will return to MFT. If one further
considers the meson emission and reabsorption (``exchange") graphs
in the baryon proper self-energy, the approximation is referred to
Hartree-Fock approximation (HF)~\cite{d,e,f}. The inclusion of the
exchange terms only makes a small correction to MFT. In fact,
after renormalization to equilibrium nuclear matter properties,
the binding energy curves in MFT and HF approximations are almost
indistinguishable~\cite{d}. This match between MFT and RHA or HF
approximations indicates that MFT is equivalent to calculate the
lowest order diagrams in the proper nucleon self-energy in field
theory. Thus one would expect to calculate higher order diagrams
to get the results beyond mean field. In~\cite{g}, through an
effective action formalism, the nuclear matter energy density was
explicitly calculated at two loop lever. It was found that the
higher order contributions were enormous, which altered the
description of the nuclear ground state qualitatively. This is
mainly because that an expansion in powers of loops is basically
an expansion in the dimensionless coupling constants which are
large in QHD. The quantum corrections are correspondingly large.
So far there are no systematic, reliable calculations of field
theory including higher order loop contributions in the study of
the nuclear matter.

When we refer to the loop expansion, we think of that in the area
of studying spontaneous symmetry breaking, the effective action
$\G(\phi)$, which is the generating functional of one particle
irreducible graphs, had been systematically calculated by summing
all the relevant Feynman graphs to a given order of the loop
expansion~\cite{h}. The effective action formalism had been
developed in several works~\cite{i,j,k,l}. A notable development
in this direction was the generalization of the effective action
for composite operators initially studied by Cornwall, Jackiw and
Tomboulis (CJT)~\cite{m}. According to this formalism the usually
effective action $\G(\phi)$ are generalized to depend not only on
$\phi(x)$, a possible expectation value of the quantum field
$\Phi(x)$, but also on $G(x,y)$, a possible expectation value of
the time-ordered product $T\Phi(x)\Phi(y)$. $G(x,y)$ is also the
propagator of the field. In this case the effective action
$\G(\phi,G)$ is the generating functional of the two particle
irreducible vacuum graphs. An effective potential $V(\phi,G)$,
which is an important theoretical tool in studying the symmetry
breaking and phase transition, can be defined by removing an
overall factor of space-time volume of the effective action.
Physical solutions demand minimization of the effective action
with respect to both $\phi$ and $G$~\cite{m}. As a result the CJT
effective potential should satisfy the stationary requirements
\bea \0{dV(\phi,G)}{d\phi}=0 \eea and \bea\label{1}
\0{dV(\phi,G)}{dG}=0. \eea This formalism was originally written
at zero temperature. Then it had been extended at finite
temperature by Amelino-Camelia and Pi where it was used for
investigations of the effective potential of the $\lambda\phi^4$
theory~\cite{n} and gauge theories~\cite{o}. It was also applied
to study the chiral symmetry breaking in effective chiral models
and QCD-like theories~\cite{p,q,r}. In studying the spontaneous
symmetry breaking system, like gauge symmetry breaking in the
electroweak system~\cite{o} and chiral symmetry breaking in the
strong interacting system~\cite{p}, $\phi$ is the order parameter
of the transition and nonzero. While in the system without
spontaneous symmetry breaking, this parameter is zero. Thus we
find the diagram expansion of the effective potential is basically
the same as that of the thermodynamic potential in a thermodynamic
system. That is to say we can use the CJT formalism to study the
thermodynamic system like the nuclear matter. In the CJT formalism
all the fields are treated as operators and it is possible to sum
a large class of ordinary perturbation-series diagrams to infinite
order that contribute to the effective potential or the
thermodynamic potential. The form of the propagators can be
consistently determined by a variational technique, as in
(\ref{1}). This resummation scheme can be used to study
non-perturbative physical effects. Substantially, the MFT or RHA
is also a certain resummation scheme. In this sense, the CJT
resummation scheme can be used to make calculations beyond MFT. In
principle the higher order contributions can be self-consistently
and clearly calculated. As well known, MFT is thermodynamical
consistent~\cite{b2}. To preserve the thermodynamical consistency
in the study of the thermodynamic system in field theory is not a
trivial problem. In~\cite{s}, the beyond mean field results was
obtained by consistently calculating the one baryon loop in the
proper meson self-energy. The results indicated a softer equation
of state and a smaller compression modulus. However, the
thermodynamical consistency had to be achieved by including
additional compensatory term which needed to be properly fixed at
zero temperature. In the CJT formalism, the thermodynamical
consistency will be automatically guaranteed by the stationary
conditions for the effective potential. However, this can be only
fulfilled if the final physical results could be self-consistently
worked out, which could be seen in our later discussion.

Our goal is to apply the CJT formalism to study the nuclear
matter. As far as we know, no such investigations have been found
in this direction as yet. In this paper, we will use the Walecka
model (also called QHD-I model)~\cite{d} and illustrate how the
thermodynamical potential can be consistently calculated in the
CJT formalism. In our calculation, the CJT thermodynamic potential
will be calculated at two-loop level with all the loop lines
treated as the full propagators, which is called Hartree-Fock
approximation in CJT formalism. However it should not be confused
to the usual HF approximation in the study of the nuclear matter.
In the CJT expansion of the thermodynamic potential, Hatree-Fock
approximation is the lowest order in coupling constant~\cite{m}.
Even in this approximation, the calculation is still very
difficult. In this paper, as the first step, we will simplify the
calculation by neglecting the medium effects to the mesons like in
the RHA or HF calculation. The thermodynamical consistency will be
achieved in our calculation. By a numerical study, we will
reproduce the binding energy curve of the nuclear matter. The
effective nucleon mass at finite temperature and the liquid-gas
phase transitions are also discussed. These results are found very
consistent with those of MFT.

The organization of the present paper is as follows. In section 2,
we formulate the CJT formalism in studying the nuclear matter
through the Walecka model. The gap equations are derived at
Hartree-Fock approximation. In section 3, we solve the gap
equations and thermodynamic potential consistently by neglecting
the medium effects to the mesons. The pressure and density can be
derived accordingly. We demonstrate how the thermodynamical
consistency is ensured by the stationary conditions in our
calculation. In section 4, we give our numerical results and make
the comparisons to the results of MFT. The last section comprises
a summary and discussion.

\section{CJT formalism in nuclear matter}
We start from the QHD-I Lagrangian which can been written as \bea
{\cal L}=\bar\p(i\g_\mu\pl^\mu-m_N+g_\s\s-g_\o\g_\mu\o^\mu)\p
+\012(\pl_\mu\s\pl^\mu\s-m_\s^2\s^2) -\014F_{\mu\nu}F^{\mu\nu}+
\012m_\o^2\o_\mu\o^\mu, \eea where $\p$, $\s$ and $\o$ are the
fields of the nucleon, sigma meson and omega meson respectively
and $F_{\mu\nu}=\pl_\mu\o_\nu-\pl_\nu\o_\mu$. According to the
Lagrangian we can write down the free propagators of the nucleon,
sigma meson and omega meson respectively as \bea
G_0(p)=\0i{/\!\!\!\!p-m_N}, \\ \Delta_0(p)=\0i{p^2-m_\s^2}, \\
D_0^{\mu\nu}(p)= \0{-ig^{\mu\nu}}{p^2-m_\o^2}. \eea

We will use the imaginary-time formalism to compute quantities at
finite temperature~\cite{s1}. Our notation is \bea \int_p
f(p)\equiv\0i{\b}\sum_n\int\0{d^3\bf p}{(2\pi)^3}f(p_0=i\o_n,{\bf
p}) , \eea where $\b$ is the inverse temperature, $\b=1/T$. We
have $\o_n=2n\pi T$ for boson and $\o_n=(2n+1)\pi T$ for fermion,
where $n=0, \pm1, \pm2, \cdots$. A baryon chemical potential $\mu$
can be introduced by replacing $p_0=i\o_n$ with $p_0=i\o_n+\mu$ in
the nucleon propagator.

\begin{figure}[tbh]
\begin{center}
\includegraphics[width=210pt,height=90pt]{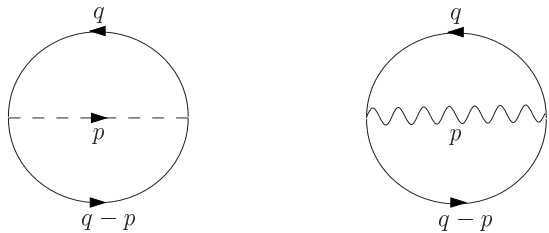}
\end{center}
\caption{Hartree-Fock approximation to $\O_2(G,\Delta,D)$. The
solid, dash and wavy lines are the nucleon propagator G, the $\s$
meson propagator $\Delta$ and the $\o$ meson propagator $D$
respectively. $q$ and $p$ are the four momenta.}\label{x1}
\end{figure}

\begin{figure}[tbh]
\begin{center}
\includegraphics[width=250pt,height=160pt]{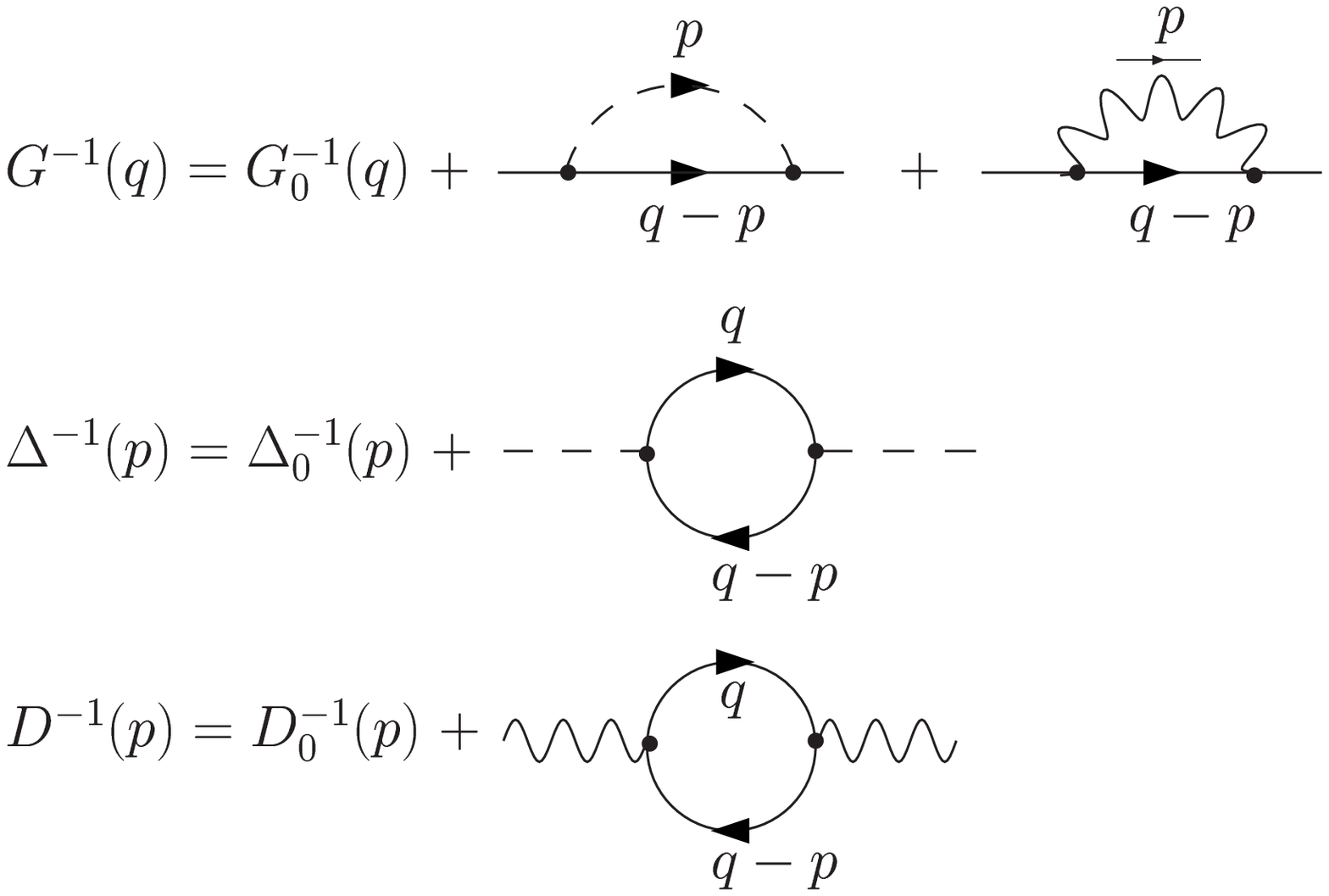}
\end{center}
\caption{The gap equations satisfied by the nucleon, the $\s$
meson and the $\o$ meson propagators at Hartree-Fock
approximation. $q$ and $p$ are the four momenta.}\label{x2}
\end{figure}

According to the CJT formalism~\cite{m}, the expansion of the
effective potential or the thermodynamic potential in the nuclear
matter can be written as \bea \O(G, \Delta, D)=i\int_p\ln
G_0(p)G^{-1}(p)&+&i\int_p\left
[G_0^{-1}(p)G(p)-1\right]-\0i2\int_p\ln\Delta_0(p)\Delta^{-1}(p)-
\0i2\int_p \left[\Delta_0^{-1}(p)\Delta(p)-1\right] \nnu
\\ &-&\0i2\int_p\ln
D_0(p)D^{-1}(p)-\0i2\int_p\left[D_0^{-1}(p)D(p)-1\right]+\O_2(G,\Delta,D),
\eea where $G$, $\Delta$ and $D$ are the full propagators of
nucleon, $\s$ meson and $\o$ meson respectively, which are
determined by the stationary condition (\ref{1}).
$\O_2(G,\Delta,D)$ is given by all the two-particle irreducible
vacuum graphs with all the propagators treated as the full
propagators. In the CJT formalism at Hartree-Fock approximation
$\O_2(G,\Delta,D)$ can be illustrated by the graphs of
Fig.\ref{x1}, where the solid lines repent $G(p)$, the dashed line
represents $\Delta(p)$ and the wavy line represents
$D^{\mu\nu}(p)$. The vertices for $\s\p\bar\p$ and
$\o^\mu\p\bar\p$ are $-g_\s$ and $g_\o\g^\mu$ respectively.
The analytic expression is \bea
\O_2(G,\Delta,D)=\0{ig_\s^2}2\int_p\int_q
Tr\left[G(q)G(q-p)\Delta(p)\right]+\0{ig_\o^2}2\int_p\int_q
Tr\left[\g^\mu G(q)\g^\nu G(q-p)D_{\mu\nu}(p)\right]. \eea From
the stationary condition (\ref{1}) which demands that
$\O(G,\Delta,D)$ be stationary against variations of $G$, $\Delta$
and $D$ respectively, we will have the following gap equations
\bea G(q)^{-1}&=&G_0(q)^{-1}+g_\s^{2}\int_{p}G(q-p)\Delta(p)+
g_\o^2\int_{p}\g^\mu G(q-p)\g^\nu D_{\mu\nu}(p), \label{2} \\
\Delta(p)^{-1}&=&\Delta_0(p)^{-1}-g_\s^2\int_q
Tr\left[G(q)G(q-p)\right], \\
D_{\mu\nu}(p)^{-1}&=&D_{0, \mu\nu}(p)^{-1}-g_\o^2\int_q
Tr\left[\g_\mu G(q)\g_\nu G(q-p)\right]. \eea The above equations
can be also represented pictorially in Fig.\ref{x2}.
These integration equations are nonlinear coupled and momentum
dependent. Needless to say they are very difficult for
computation. The certain approximations should be adopted.

\section{solving gap equations and thermodynamic potential}
In this section we will solve the gap equations through certain
approximations. The thermodynamic potential will be consistently
determined. Then we will derive the pressure and the net baryon
density and demonstrate how the thermodynamical consistency is
achieved.

To solve the equations, we need first to decouple the equations.
As in usual Hartree or Hartree-Fock approximation in studying the
nuclear matter, the meson propagators are treated as the free
propagators~\cite{d}. In this paper, as the first step, we will
simplify the equations by replacing the meson propagators by the
free ones. This also means we have neglected the medium effects to
the mesons. We expect that under this approximation it could yield
the results which are consistent with those of the MFT. The
approximation means \bea\label{2a} \Delta(p)=\Delta_0(p), \ \ \ \
D_{\mu\nu}(p)=D_{0,\mu\nu}(p). \eea As a result the thermodynamic
potential is reduced to \bea\label{2b} \O(G, \Delta_0,
D_0)=&i&\int_p\ln
G_0(p)G^{-1}(p)+i\int_p\left[G_0^{-1}(p)G(p)-1\right] \nnu
\\ &+&\0{ig_\s^2}2\int_p\int_q
Tr\left[G(q)G(q-p)\Delta_0(p)\right]+\0{ig_\o^2}2\int_p\int_q
Tr\left[\g^\mu G(q)\g^\nu G(q-p)D_{0,\mu\nu}(p)\right]. \eea The
gap equations are much simplified accordingly. Only the nucleon
gap equation (\ref{2}) is left and needs to be solved. To proceed
we take the following ansatz of the full nucleon propagator
\bea\label{3} G(q)=\0i{\g_{\mu}q^{\mu}-m_N+\S}, \eea where $\S$ is
the proper nucleon self-energy which will be determined by the
equation (\ref{2}). It can be generally written as \bea\label{4}
\S(q)=\S_s(q)-\g^0\S_0(q)+{\bm\g}{\bm\cdot}{\bf q}\S_v(q). \eea
Thus we can define an effective nucleon mass \bea\label{4a}
M_N(q)=m_N-\S_s(q), \eea which is momentum dependent. As the
nuclear matter is a uniform system at rest, the $\S_v$ term in
equation (\ref{4}) is usually neglected as in RHA
calculation~\cite{c}. In the usual HF approximation in the study
of the nuclear matter, the contribution of the $\S_v$ term to the
final physical result is also found to be very small~\cite{d}. In
our discussion this term will be neglected. Substituting equations
(\ref{3}) and (\ref{4}) into equation (\ref{2}), after some
calculations on separating the terms which are associated with and
without $\g$ matrix, and then by comparing the two sides of the
equation, we can obtain the following equations for the components
of $\S$ \bea
\S_s(q)=ig_\s^2\int_p\0{-M_N(q)}{\left[(q_0^*-p_0)^2-E_{\bf
q-p}^2\right]\left[p_0^2-E_\s({\bf p})^2\right]}
+ig_\o^2\int_p\0{4M_N(q)}{\left[(q_0^*-p_0)^2-
E_{\bf q-p}^2\right]\left[p_0^2-E_\o({\bf p})^2\right]}, \\
\S_0(q)=ig_\s^2\int_p\0{q_0^*-p_0}{\left[(q_0^*-p_0)^2-E_{\bf
q-p}^2\right]\left[p_0^2-E_\s({\bf
p})^2\right]}+ig_\o^2\int_p\0{2q_0^*-2p_0}{\left[(q_0^*-p_0)^2-
E_{\bf q-p}^2\right]\left[p_0^2-E_\o({\bf p})^2\right]}, \eea
where $q_0^*=q_0-\S_0(q)$, $E_{\s/\o}({\bf p})=\sqrt{{\bf
p^2}+m_{\s/\o}^2}$ and $E_{\bf q-p}=\sqrt{({\bf
q-p})^2+M_N(q)^2}$. After we perform the sums over the Matsubara
frequencies, we obtain \bea \S_s(q)&=&-g_\s^2\int \0{d^3\bf
p}{(2\pi)^3} \0{M_N(q)}{2E_\s({\bf p}) E_{\bf q-p}}
\left[A_\s-B_\s\right]+g_\o^2\int\0{d^3\bf
p}{(2\pi)^3}\0{2M_N(q)}{E_\o({\bf p}) E_{\bf
q-p}}\left[A_\o-B_\o\right], \label{5}
\\ \S_0(q)&=&g_\s^2\int \0{d^3\bf
p}{(2\pi)^3} \01{2E_\s({\bf p}) E_{\bf q-p}}
\left[C_\s-D_\s\right]+g_\o^2\int\0{d^3\bf
p}{(2\pi)^3}\01{E_\o({\bf p}) E_{\bf q-p}}\left[C_\o-D_\o\right],
\label{6} \eea where the expressions of $A_i$, $B_i$, $C_i$ and
$D_i$ ($i=\s,\o$) are given in the following, \bea
A_i=\0{n_i(q_0^*-E_{\bf q-p})-\tilde{n}_+ E_i}{(q_0^*-E_{\bf
q-p})^2-E_i^2}, \ \ && \ \ B_i=\0{n_i(q_0^*+E_{\bf
q-p})+\tilde{n}_- E_i}{(q_0^*+E_{\bf q-p})^2-E_i^2}, \\
C_i=\0{q_0^*(q_0^*-E_{\bf q-p})n_i-E_i^2n_i- E_iE_{\bf
q-p}\tilde{n}_+}{(q_0^*-E_{\bf q-p})^2-E_i^2}, \ \ && \ \
D_i=\0{q_0^*(q_0^*+E_{\bf q-p})n_i-E_i^2n_i- E_iE_{\bf
q-p}\tilde{n}_-}{(q_0^*+E_{\bf q-p})^2-E_i^2}, \eea in which
$E_i=\sqrt{{\bf p}^2+m_i^2}$ and \bea n_i=\01{e^{\b E_i}-1}, \ \ \
\ \ \ \tilde n_\pm=\01{e^{\b(E_{\bf q-p}\mp\mu^*)}+1}, \eea where
$\mu^*=\mu-\S_0(q)$ and $\mu$ is the baryon chemical potential. In
obtaining equations (\ref{5}) and (\ref{6}), we have encountered
the divergent terms which are independent of the distribution
functions $n_i$ and $\tilde n_\pm$. In principle these divergent
parts could be properly renormalized. However, in this rudimentary
work, we simply neglect the divergent parts which are not explicit
temperature dependent. This treatment is not strange in the
literature~\cite{p,u}.

Now the equations are finite but momentum dependent, which are
still intractable. We note that in studying the nuclear matter,
the effective mass is usually defined as the pole of the full
propagator in the limit ${\bf q}\to 0$~\cite{v,t}. In light of
this definition, the effective nucleon mass here can be defined by
\bea\label{6m} \left.{q_0^*}^2-{\bf q}^2-M_N^2(|q_0^*|=M_N,|{\bf
q}|=0)\right|_{|{\bf q}|=0}=0. \eea This means in equations
(\ref{5}) and (\ref{6}) we will take $|{\bf q}|=0$ and set
$|q_0^*|=M_N$. The treatment is different from the usual pole
approximation for that we will take $|{\bf q}|=0$ before the
angular integration performed. This will greatly simplify the
calculation. Furthermore, as $|q_0^*|=M_N$, $q_0^*$ can be set to
either positive or negative value. In order to ensure that at
$\mu=0$ the baryon density keeps zero which can be realized in our
later discussion, here we will take $q_0^*=M_N$ in $A_i$ and
$B_i$, while take $q_0^*=-M_N$ in $C_i$ and $D_i$. After these
approximations and treatments, the equations (\ref{5}) and
(\ref{6}) will be further simplified to \bea\label{5a}
\S_s=-g_\s^2\int \0{d^3\bf p}{(2\pi)^3} \0{M_N}{2E_\s
E}\0{2(M_N-E)n_\s-E_\s(\tilde n_++\tilde
n_-)}{(M_N-E)^2-E_\s^2}+g_\o^2\int \0{d^3\bf p}{(2\pi)^3}
\0{2M_N}{E_\o E}\0{2(M_N-E)n_\o-E_\o(\tilde n_++\tilde
n_-)}{(M_N-E)^2-E_\o^2}, \eea \bea\label{6a} \S_0=g_\s^2\int
\0{d^3\bf p}{(2\pi)^3} \0{-(\tilde n_+-\tilde n_-)}
{2(M_N-E)^2-2E_\s^2}+g_\o^2\int \0{d^3\bf p}{(2\pi)^3} \0{-(\tilde
n_+-\tilde n_-)}{(M_N-E)^2-E_\o^2}, \eea where $E=\sqrt{{\bf
p}^2+M_N^2}$. Now $\S_s$, $\S_0$ and $M_N$ become momentum
independent and can be numerically calculated.

In the next we will demonstrate how the thermodynamic potential,
the pressure and net baryon density can be derived in a
thermodynamical consistent way.

Considering equation (\ref{2}), (\ref{2a}) and (\ref{3}) we can
have \bea G_0(q)^{-1}G(q)-1=i\S(q) G(q), \eea \bea\label{6b}
\S(q)=ig_\s^2\int_pG(q-p)\Delta_0(p)+ig_\o^2\int_p\g_\mu
G(q-p)\g_\nu D_0^{\mu\nu}(p). \eea Substituting the above
equations into equation (\ref{2b}), the thermodynamic potential
will be reduced to \bea\label{7} \O=i\int_p\ln
G_0(p)G(p)^{-1}-\012\int_qTr\left[G(q)\S(q)\right]. \eea The
pressure can be obtained by the thermodynamic relation
\bea\label{8} P=-\O. \eea For the consistency of the calculation
under our approximation, $\S(q)$ in equation (\ref{7}) will be
also determined at $|{\bf q}|=0$ and $|q_0^*|=M_N$, which means
$\S$ becomes independent of the momentum in equation (\ref{7}).
After the frequency sums in equation (\ref{7}) and considering
equation (\ref{8}) we can obtain the pressure as \bea\label{9}
\lefteqn{\hspace{-3cm}P(\mu,T)=\043\int \0{d^3\bf
p}{(2\pi)^3}\0{\bf{p}^2}E\left[\01{e^{\b(E-\mu^*)}+1} +
\01{e^{\b(E+\mu^*)}+1}\right] -2M_N\S_s\int \0{d^3\bf
p}{(2\pi)^3}\01E\left[\01{e^{\b(E-\mu^*)}+1}+\01{e^{\b(E+\mu^*)}+1}\right]}
\hspace{3cm} \nnu \\ &&+2\S_0 \int \0{d^3\bf
p}{(2\pi)^3}\left[\01{e^{\b(E-\mu^*)}+1}-\01{e^{\b(E+\mu^*)}+1}\right],
\eea where $\S_s$ and $\S_0$ are determined by solving equations
(\ref{5a}) and (\ref{6a}). In (\ref{9}) we have also neglected the
divergent terms which are independent of the distribution
functions.

The thermodynamical consistency requires that at fixed $T$ and
$\mu$ the pressure will be maximized with respect to $\S_s$ or
$\S_0$, when $\S_s$ and $\S_0$ are independent variables. This
requirement will be satisfied by the stationary condition in the
CJT formalism. To make it clear, we assume that $P$ is still in
its functional form of $G$, then from equation (\ref{3}), which
indicates $G$ is a function of $\S$, and considering the
stationary condition (\ref{1}) we will have \bea\label{10}
\0{dP(G(\S))}{d\S}=\0{dP(G(\S))}{dG}\0{dG(\S)}{d\S}=0. \eea This
equation ensures the thermodynamical consistency. So when the
pressure or the thermodynamic potential is self-consistently
determined by solving the gap equations, which are derived by the
stationary conditions, the thermodynamical consistency will be
automatically achieved. However, one should notice that the
pressure in the form of equation (\ref{9}) can not be maximized
with respect to $\S_s$ or $\S_0$, because in obtaining equation
(\ref{9}), the gap equations have been already substituted into
it, which makes that $\S_s$ and $\S_0$ are not independent
variables in equation (\ref{9}).

Next we will derive the net baryon density of the system. The
density will be determined by the thermodynamic relation \bea
\r=\left.\0{\pl P}{\pl\mu}\right|_T. \eea From a general
expression of the pressure, which is a function of the independent
variables $T$, $\mu$ and $\S$, we can write down the partial
derivative in the following expression \bea\label{11} \left.\0{\pl
P(\S,\mu,T)}{\pl\mu}\right|_T=\left.\0{\pl
P(\S,\mu,T)}{\pl\mu}\right|_{\S,T}+ \left.\0{\pl
P(\S,\mu,T)}{\pl\S}\right|_{\mu,T}\left.\0{\pl\S}{\pl\mu}\right|_T.
\eea According to equation (\ref{10}) the second term on the r.h.s
of equation (\ref{11}) becomes to zero, which shows a fulfillment
of the thermodynamical consistency. The partial derivative is
reduced to \bea \left.\0{\pl
P(\S,\mu,T)}{\pl\mu}\right|_T=\left.\0{\pl
P(\S,\mu,T)}{\pl\mu}\right|_{\S,T}. \eea Then from equation
(\ref{2b}) and considering equation (\ref{8}), the derivative can
be formally evaluated as \bea\label{12} \left.\0{\pl
P(\S,\mu,T)}{\pl\mu}\right|_T&=&-i\0{\pl}{\pl\mu}\int_p\ln\left[G_0(p)G(p)^{-1}\right]
+\int_qTr\left[\0{\pl G(q)}{\pl\mu}\S(q)\right] \nnu \\
&-&\int_qTr\left\{\0{\pl
G(q)}{\pl\mu}\left[ig_\s^2\int_pG(q-p)\Delta_0(p)+ig_\o^2\int_p\g_\mu
G(q-p)\g_\nu D_0^{\mu\nu}(p)\right]\right\}. \eea From equation
(\ref{6b}) we find that the second term and third term on r.h.s of
equation (\ref{12}) will cancel each other, thus the final result
is \bea \left.\0{\pl P(\S,\mu,T)}{\pl\mu}\right|_T=
-i\0{\pl}{\pl\mu}\int_p\ln\left[G_0(p)G(p)^{-1}\right]. \eea Then
the net baryon density is given as \bea\label{13} \r=4\int\0{d^3
\bf
p}{(2\pi)^3}\left[\01{e^{\b(E-\mu^*)}+1}-\01{e^{\b(E+\mu^*)}+1}
\right]. \eea This is the standard form of net density for
quasi-particles, which also indicates that the thermodynamic
functions could be calculated thermodynamical consistently.

Furthermore the energy density of the system can be derived by
\bea\label{14} \e=-P+\mu\r+T\0{\pl P}{\pl T}. \eea In the
following we can study the thermodynamics of the nuclear matter.

\section{numerical results and comparison to MFT}
\begin{figure}[tbh]
\begin{center}
\includegraphics[width=220pt,height=140pt]{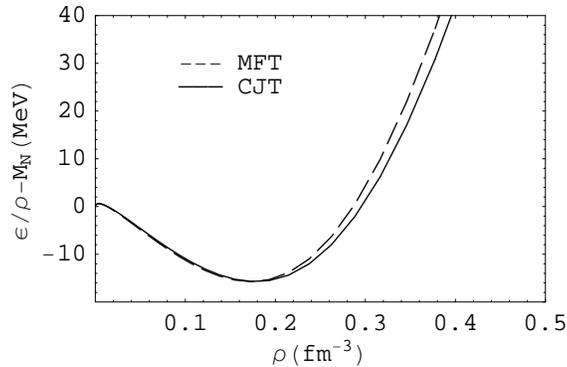}
\end{center}
\caption{The average energy per nucleon minus the nucleon mass
$m_N$ as functions of baryon density at zero temperature in the
CJT formalism (solid line) and MFT (dashed line).}\label{x3}
\end{figure}

The coupling constants $g_\s$ and $g_\o$ will be refitted by
reproducing the correct saturation properties of the nuclear
matter at zero temperature. The nucleon mass, sigma meson mass and
omega meson mass are taken as $m_N=939MeV$, $m_\s=550MeV$ and
$m_\o=783MeV$. By setting $T=0$, from equation (\ref{9}),
(\ref{13}) and (\ref{14}) we can reproduce the correct binding
energy curve of the nuclear matter which means at a saturation
density of $0.16$ nucleons per $fm^3$ it has a binding energy of
$16MeV$ per nucleon. The curve of energy per nucleon versus
density is shown in Fig.\ref{x3}. The coupling constants thus are
fixed at $g_\s^2=155.6$ and $g_\o^2=521.5$. These values look a
little greater than those of the MFT which are $g_\s^2=91.6$ and
$g_\o^2=136.2$~\cite{d}.
\begin{figure}[tbh]
\begin{center}
\includegraphics[width=220pt,height=140pt]{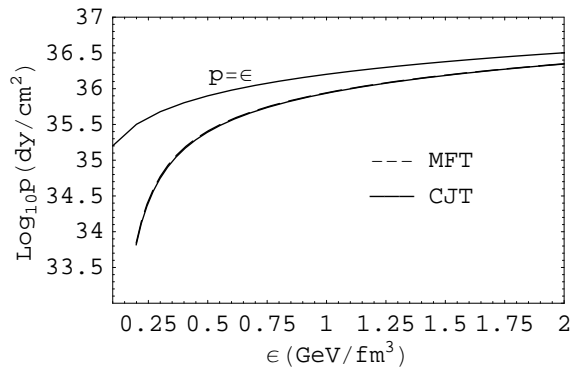}
\end{center}
\caption{The nuclear matter equations of state at zero temperature
in the CJT formalism (solid line) and MFT (dashed
line).}\label{x4}
\end{figure}
The binding energy curve of MFT is also plotted in Fig.\ref{x3} in
dashed line. We can see that the result of CJT is very close to
that of MFT. At low density the energy curve of CJT is almost
overlap with the energy curve of MFT; at high density the curve of
CJT rises a little slower than that of MFT. The compressibility of
nuclear matter at saturation density calculated in the CJT
formalism turns out to be \bea K\equiv
p_F^2\0{d^2(\e/\r)}{dp_F^2}=485MeV, \eea where $p_F$ is the Fermi
momentum. It is smaller compared to that of the MFT which is
$K=540MeV$. The equation of state $P$ vs. $\e$ for nuclear matter
is shown in Fig.\ref{x4}. The CJT and MFT results are almost
overlap. Note the approach from below to the casual limit $\e=P$
(where $c_{sound}=c_{light})$ at high density.

\begin{figure}[tbh]
\begin{center}
\includegraphics[width=220pt,height=140pt]{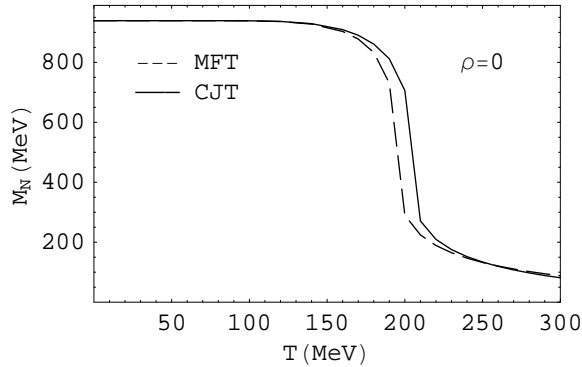}
\end{center}
\caption{The effective nucleon mass as functions of temperature at
zero density in the CJT formalism (solid line) and MFT (dashed
line).}\label{x5}
\end{figure}
\begin{figure}[tbh]
\begin{center}
\includegraphics[width=220pt,height=140pt]{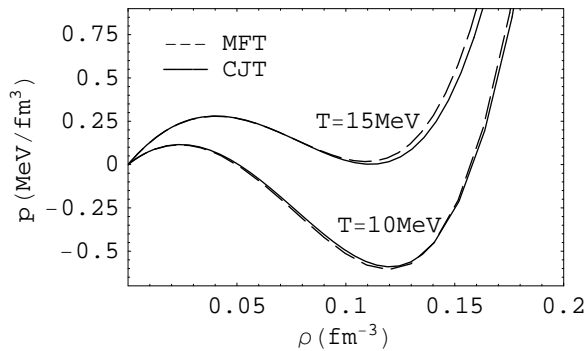}
\end{center}
\caption{The isotherms of $P$ vs. $\r$ at $T=10MeV$ and $T=15MeV$
in the CJT formalism (solid line) and MFT (dashed
line).}\label{x6}
\end{figure}

At finite temperature, from equation (\ref{4a}) and considering
(\ref{6m}), the effective nucleon mass as a function of
temperature at zero density can be plotted in Fig.\ref{x5} (solid
line). When compared to that of MFT (dashed line), the effective
mass from the CJT calculation drops a little more slowly at high
temperature. The well-known liquid-gas phase transition of the
nuclear matter at low temperature also exists by the CJT
calculation. The isotherms of $P$ vs. $\r$, which show the first
order transitions, are plotted in Fig.\ref{x6}. The CJT results
are again found very close to the MFT results. The critical
temperature in the CJT calculation is about $19MeV$ which is
almost the same as that in the MFT.

\section{summary}
In this paper we have made an attempt to use the CJT formalism to
study the nuclear matter. As the first step, we have neglected the
medium effects to the mesons and obtained the results which are
found very consistent with those from MFT. We have also
demonstrated how the thermodynamical consistency has been achieved
in the CJT formalism in studying the nuclear matter. In our
discussion one can also see that the beyond mean field calculation
of the nuclear matter can be carried out in the CJT formalism, at
least by including the medium effects to the mesons. However it is
obvious that the calculation will be quite involved. This topic
will be studied in our further work.

\begin{acknowledgments}
This work was supported in part by the National Natural Science
Foundation of China with No. 90303007 and the Ministry of
Education of China with Project No. 704035.
\end{acknowledgments}

\end{document}